\documentclass[nonblindrev,copyedit]{informs3} 

\usepackage{graphicx}
\usepackage{setspace,graphics,booktabs,color,enumerate,subfigure,url}
\usepackage{amssymb}
\usepackage{mathrsfs}
\usepackage{multirow}
\usepackage{amsmath, amssymb, color}

\usepackage{subfigure}
\definecolor{DarkBlue}{rgb}{0,0.08,0.45}
\usepackage[backref = false, bookmarks, colorlinks = true, plainpages = false, citecolor = DarkBlue , urlcolor = DarkBlue, filecolor = DarkBlue, linkcolor = DarkBlue]{hyperref}
\usepackage{array}
\usepackage{float}
\floatstyle{ruled}
\newfloat{algorithm}{htp}{lop}
\floatname{algorithm}{Algorithm}

\newtheorem{definition}{Definition}

\newcommand{\bx}{\mbox{\boldmath $x$}}
\newcommand{\bu}{\mbox{\boldmath $u$}}
\newcommand{\by}{\mbox{\boldmath $y$}}
\newcommand{\bv}{\mbox{\boldmath $v$}}
\newcommand{\bdb}{\mbox{\boldmath $b$}}
\newcommand{\bp}{\mbox{\boldmath $p$}}
\newcommand{\bq}{\mbox{\boldmath $q$}}
\newcommand{\bw}{\mbox{\boldmath $w$}}

\newcommand{\OPT}{{\mbox{OPT}}}

\usepackage{natbib}
 \bibpunct[, ]{(}{)}{,}{a}{}{,}%
 \def\bibfont{\small}%
\theoremstyle{TH}%
\newtheorem{theorem}{Theorem}
\newtheorem{lemma}{Lemma}
\newtheorem{proposition}{Proposition}

\newtheorem{assumption}{Assumption}

\ECRepeatTheorems

\EquationsNumberedThrough    

\MANUSCRIPTNO{}

\begin{document}




\TITLE{\large{A Dynamic Learning Algorithm for Online Matching
Problems with Concave Returns}}


\ARTICLEAUTHORS{ \AUTHOR {Xiao Alison Chen} \AFF{Department of
Industrial and Systems Engineering, University of Minnesota, MN, USA
\href{mailto:chen2847@umn.edu}{chen2847@umn.edu}}\AUTHOR {Zizhuo
Wang} \AFF{Department of Industrial and Systems Engineering,
University of Minnesota, MN, USA
\href{mailto:zwang@umn.edu}{zwang@umn.edu}} }

\date{\today}

\ABSTRACT {We consider an online matching problem with concave
returns. This problem is a generalization of the traditional online
matching problem and has vast applications in online advertising. In
this work, we propose a dynamic learning algorithm that achieves
near-optimal performance for this problem when the inputs arrive in
a random order and satisfy certain conditions. The key idea of our
algorithm is to learn the input data pattern dynamically: we solve a
sequence of carefully chosen partial allocation problems and use
their optimal solutions to assist with the future decisions. Our
analysis belongs to the primal-dual paradigm; however, the absence
of linearity of the objective function and the dynamic feature of
the algorithm makes our analysis quite unique. We also show through
numerical experiments that our algorithm performs well for test
data.}


\maketitle

\section{Introduction}
\label{sec:intro}

In traditional optimization models, inputs are usually assumed to be
known and efficient algorithms are sought to find the optimal
solutions. However, in many practical cases, data does not reveal
itself at the beginning. Instead, it comes in an online fashion. For
example, in many revenue management problems, customers arrive
sequentially and each time a customer arrives, the decision maker
has to make some irrevocable decisions (e.g., what product to sell,
at what prices) for this customer without knowing any of the future
inputs. Such a regime is often called {\it online optimization}.
Online optimization has gained much attention in the research
community in the past few decades due to its applicability in many
practical problems, and much effort has been directed toward
understanding the quality of solutions that can be obtained under
such settings. For an overview of the online optimization literature
and its recent developments, we refer the readers to
\citet{borodin}, \citet{buchbinder-or09} and \citet{Devanur2011}.

In this paper, we consider a special type of online optimization
problem - an online matching problem. Online matching problems are
considered as fundamental problems in online optimization theory and
have important applications in the online advertisement allocation
problems. For a review of online matching problems, we refer the
readers to \citet{mehta2012}.
In the problem we study, there is an underlying weighted bipartite
graph $G = (I, J, E)$ with weights $b_{ij}$ for each edge $(i,j) \in
E$. The vertices in $J$ arrive sequentially in some order, and
whenever a vertex $j\in J$ arrives, the set of weights $b_{ij}$ is
revealed for all $i\in I$, $(i,j)\in E$. The decision maker then has
to match $j$ to one of its neighbors $i$, and a value of $b_{ij}$
will be obtained from this matching. In our problem, the decision
maker's gain from each vertex $i$ is a function of the total matched
value to this vertex, and his goal is to maximize the total gain
from
all vertices. 
Mathematically, the problem can be formulated as follows
(assume $|I| = m$, $|J| = n$, and let $b_{ij} = 0$ for $(i,j) \notin E$):
\begin{equation}\label{offlineconcave}
\begin{array}{lll}
\mbox{maximize}_{\bx} & \sum_{i=1}^mM_i\left(\sum_{j=1}^n
b_{ij}x_{ij}\right)\\
\mbox{s.t.} & \sum_{i=1}^m x_{ij} \le 1, & \forall j \\
& x_{ij} \ge 0, & \forall i,j,
\end{array}
\end{equation}
where $x_{ij}$ denotes the fraction of vertex $j$ that is matched to
vertex $i$.\footnote{We allow fractional allocations in our model.
However, our proposed algorithms output integer solutions. Thus all
our results hold if one confines to integer solutions.} In
(\ref{offlineconcave}), the coefficient $\bdb_j =
\{b_{ij}\}_{i=1}^m$ is revealed only when vertex $j$ arrives, and an
irrevocable decision $\bx_j = \{x_{ij}\}_{i=1}^m$ has to be made
before observing the next input. For each $i$, $M_i(\cdot)$ is a
nondecreasing concave function with $M_i(0) = 0$. In this paper, we
assume that $M_i(\cdot)$s are continuously differentiable.

As mentioned earlier, online matching problems have a very important
application in the online advertisement allocation problem, which we
will later refer to as the Adwords problem. In the Adwords problem,
there are $m$ advertisers (which we also call the bidders). A
sequence of $n$ keywords are searched during a fixed time horizon.
Based on the relevance of the keyword, the $i$th bidder would bid a
certain amount $b_{ij}$ to show his advertisement on the result page
of the $j$th keyword. The search engine's decision is to allocate
each keyword to one of the $m$ bidders (we only consider a single
allocation in this paper). Note that each allocation decision can
only depend on the information earlier in the arrival sequence but
not on any future data. As pointed out in \citet{Devanur_concave},
there are several practical motivations for considering a concave
function of the matched bids in the Adwords problem. Among them are
convex penalty costs for under-delivery in search engine-advertiser
contracts, the concavity of the click-through rate in the number of
allocated bids observed in empirical data and fairness
considerations. In each of the situations mentioned above, one can
write the objective as a concave function. We refer the readers to
\citet{Devanur_concave} for a more thorough review of the
motivations for this problem. It is worth noting that there is a
special case of this problem where $M_i(x) = \min\{x, B_i\}$. In
this case, one can view that the bidder has a budget $B_i$ and the
revenue from each bidder is bounded by $B_i$.

%


One important question when studying online algorithms is the
assumptions on the input data. In this work, we adopt a {\it random
permutation model}. More precisely, we assume:
\begin{enumerate}
\item The total number of arrivals $n = |J|$ is known a priori.
\item The weights $\{b_{ij}\}$ can be adversarially
chosen. However, the order that $j$ arrives is uniformly distributed
over all the permutations.
\end{enumerate}
The random permutation model has been adopted in much recent
literature in the study of online matching problems, see, e.g.,
\citet{Devanur, feldman2010online, wang}, etc. It is equivalent to
saying that a set of $\mathcal{B} = \{\tilde{\bdb}_1,
\tilde{\bdb}_2, ...,\tilde{\bdb}_n\}$ is arbitrarily chosen
beforehand (unknown to the decision maker). Then the arrivals
$\bdb_1,\bdb_2,...,\bdb_n$ are drawn randomly without replacement
from $\mathcal{B}$. The random permutation model is an intermediate
path between using a worst-case analysis and assuming each input
data is drawn independently and identically distributed (i.i.d.)
from a certain distribution. On one hand, compared to the worst-case
analysis (see, e.g., \citealt{mehta, buchbinder07esa, feldman2009,
Devanur_concave}), the random permutation model is practically
reasonable yet much less conservative. On the other hand, the random
permutation model is much less restrictive than assuming the inputs
are drawn i.i.d. from a certain distribution
(\citealt{Devanur2011}). Also, the assumption of the knowledge of
$n$ is necessary for any online algorithm to achieve near-optimal
performance (see \citealt{Devanur}). Therefore, for large problems
with relatively stationary inputs, the random permutation model is a
good approximation and the study of such models is of practical
interest. Next we define the performance measure of an algorithm
under the random permutation model:
\begin{definition}[$c$-competitiveness]
Let $\OPT$ be the optimal value for the offline problem
(\ref{offlineconcave}). An online algorithm $A$ is called
$c$-competitive in the random permutation model if the expected
value of the online solutions by using $A$ is at least $c$ times the
optimal value of (\ref{offlineconcave}), that is
\begin{eqnarray*}
{\mathbb E}_\sigma \left[ \sum_{i=1}^m M_i\left(\sum_{j=1}^n
b_{ij}x_{ij}(\sigma, A)\right)\right] \ge c \OPT,
\end{eqnarray*}
where the expectation is taken over uniformly random permutations
$\sigma$ of $1,...,n$, and $x_{ij}(\sigma, A)$ is the $ij$th
decision made by algorithm $A$ when the inputs arrive in order
$\sigma$.
\end{definition}

In \citet{Devanur_concave}, the authors propose an algorithm for the
online matching problem with concave returns that has a constant
competitive ratio under the {\it worst-case model} (the constant
depends on the forms of each $M_i(\cdot)$). They also show that a
constant competitive ratio is the {\it best} possible result under
that model. In this paper, we propose an algorithm under the random
permutation model, which achieves {\it near-optimal performance}
under some conditions on the input.

Our main result is stated as follows:
\begin{theorem}\label{thm:main}
Fix $\epsilon \in (0, 1/2)$. There exists an algorithm (Algorithm
DLA) that is $1-\epsilon$ competitive for the online matching
problem with concave returns $M_i(\cdot)$s under the random
permutation model if
\begin{eqnarray}\label{condition_n}
n \ge \Omega \left( \max\left\{\frac{ \log{(m/\epsilon)}}{\epsilon
\bar{b}^2},
\frac{m^2\log{(m^2n/\epsilon)}F(M,\eta)}{\epsilon^3\bar{b}}\right\}
\right),
\end{eqnarray}
where $\bar{b} = \frac{1}{n}\min_i\{\sum_{j=1}^nb_{ij}\}$, $\eta =
\frac{\min_{i,j}\{b_{ij}|b_{ij} > 0\}}{\max_{i,j} b_{ij}}$, and
$F(M, \eta)$ is a constant that only depends on each $M_i(\cdot)$
and $\eta$.
\end{theorem}
In condition (\ref{condition_n}), $\bar{b}$ can be viewed as the
average bid value of a bidder over time. Given that each bidder is
at least interested in some fractions of the keywords, this average
will go to a certain constant as $n$ becomes large. Also, $\eta$ can
be viewed as the ratio between the value of the smallest non-zero
bid and the highest bid. In practice, this is often bounded below by
a constant by enforcing a reserve price and a maximum price for any
single bid. The exact functional form of $F(M, \eta)$ is somewhat
complicated, and is given in Proposition \ref{prop:ola_condition1}.
Just to give an example, if we choose $M_i(x) = x^p$ ($0 < p <1$),
then $F(M, \eta) = \frac{2}{\eta^{(2-p)/(1-p)}}$. Therefore,
condition (\ref{condition_n}) can be viewed as simply requiring the
total number of inputs is large, which is often the case in
practice. For example, in the Adwords problem, $n$ is the number of
keyword searches in a certain period, and for instance, Google
receives more than $5$ billion searches per day. Even if we focus on
a specific category, the number can still be in the millions. Thus,
this condition is reasonable. We note that most learning algorithms
in the literature make similar requirements, see \citet{Devanur,
wang}, and \citet{Molinaro2012}. Furthermore, as we will show in our
numerical tests, our algorithm performs well even for problems with
sizes that are significantly smaller than the condition requires,
which validates the potential usefulness of our algorithm.

%
%

To propose an algorithm that achieves near-optimal performance, the
main idea is to utilize the observed data in the allocation process.
In particular, since the input data arrives in a random order, using
the past input data and projecting it into the future should present
a good approximation for the problem. To mathematically capture this
idea, we use a primal-dual approach. We obtain the dual optimal
solutions to suitably constructed optimization problems and use them
to assist with future allocations. We first propose a one-time
learning algorithm (OLA, see Section \ref{sec:ola}) that only solves
an optimization problem once at time $\epsilon n$. By carefully
examining this algorithm, we prove that it achieves near-optimal
performance when the inputs satisfy certain conditions. However, the
conditions are stronger than those stated in Theorem \ref{thm:main}.
To improve our algorithm, we further propose a dynamic learning
algorithm (DLA, see Section \ref{sec:dla}). The dynamic learning
algorithm makes better use of the observed data and updates the dual
solution at a geometric pace, that is, at time $\epsilon n$,
$2\epsilon n$, $4\epsilon n$ and so on. We show that these
resolvings can lift the performance of the algorithm and thus prove
Theorem \ref{thm:main}. As one will see in the proof of the DLA, the
choice of the resolving points perfectly balances the trade-off
between {\it exploration} and {\it exploitation}, which are the main
trade-offs in such types of learning algorithms.

It is worth mentioning that a similar kind of dynamic learning
algorithm has been proposed in \citet{wang} and further studied in
\citet{wang_thesis} and \citet{Molinaro2012}. However, those works
only focus on linear objectives. In our analysis, the nonlinearity
of the objective function presents a non-trivial hurdle since one
can no longer simply analyze the revenue generated in each time
segment and add them together. In this paper, we successfully work
around this hurdle by a convex duality argument. We believe that our
analysis is a non-trivial extension of the previous work. Moreover, the
problem solved has important applications.

The remainder of the paper is organized as follows. In Section
\ref{sec:ola}, we start with a one-time learning algorithm and prove
that it achieves near-optimal performance under some mild conditions
on the input. The one-time learning algorithm is easy to understand
and shows important insights for designing this class of learning
algorithms. However, it only achieves a weaker performance than what
is stated in Theorem \ref{thm:main}. In Section \ref{sec:dla}, we
propose a dynamic learning algorithm which makes better use of the
data and has a stronger performance. Some numerical test results of
our algorithm are presented in Section \ref{sec:numerical}, which
validate the strength of our algorithm. Section \ref{sec:conclusion}
concludes this paper.

\section{One-Time Learning Algorithm}
\label{sec:ola} We first rewrite the offline problem
(\ref{offlineconcave}) as follows:
\begin{equation}
\begin{array}{lll}\label{primal}
\mbox{maximize}_{\bx, \bu} & \sum_{i=1}^m M_i(u_i) &\\
\mbox{s.t.} & \sum_{j=1}^n b_{ij}x_{ij} = u_i, & \forall i \\
& \sum_{i=1}^m x_{ij} \le 1, & \forall j \\
& x_{ij}\ge 0, & \forall i,j.
\end{array}
\end{equation}
We define the following {\it dual problem}:
\begin{equation}
\begin{array}{lll}\label{dual} \mbox{inf}_{\bv, \by} & \sum_{j=1}^n y_j + \sum_{i=1}^m
\left(M_i(v_i) -
M_i'(v_i)v_i\right)\\
\mbox{s.t.}& y_j \ge b_{ij}M_i'(v_i),\quad\quad \forall i,j\\
& v_i\ge 0, \quad\quad\forall i\\
& y_j\ge 0, \quad\quad\forall j.
\end{array}
\end{equation}
Let the optimal value of (\ref{primal}) be $P^*$ and the optimal
value of (\ref{dual}) be $D^*$. In \citet{Devanur_concave}, the
authors proved the weak duality between (\ref{primal}) and
(\ref{dual}). In the following lemma, we prove that in fact the
strong duality holds. The proof of the lemma is relegated to the
Appendix.
\begin{lemma}\label{lem:strongduality}
$P^* = D^*$. Furthermore, the objective value of any feasible
solution to (\ref{dual}) is an upper bound of $P^*$.
\end{lemma}
%
Before we describe our algorithm, we define the following partial
optimization problem:
\begin{equation}
\begin{array}{rlll}\label{partial}
(\bf P_{\epsilon}) & \mbox{maximize}_{\bx, \bu} & \sum_{i=1}^m M_i(u_i)\\
&\mbox{s.t.} & \sum_{j=1}^{\epsilon n} \frac{b_{ij}}{\epsilon}
x_{ij}
= u_i, &\forall i\\
&& \sum_{i=1}^m x_{ij} \le 1, & \forall j\\
&& x_{ij} \ge 0, & \forall i,j.
\end{array}
\end{equation}
Now we define the one-time learning algorithm as follows:
\begin{algorithm}
\caption{One-Time Learning Algorithm
(OLA)}\label{alg:onetimelearning}
\begin{enumerate}
\item During the first $\epsilon n$ arrivals, no allocation is made.
\item After observing the first $\epsilon n$ arrivals,
solve ($\bf P_{\epsilon}$) and denote the optimal solutions by
$\hat{\bx}$ and $\hat{\bu}$.
\item For any $m$ dimensional vector $\bw, \bq\ge 0$, define
\begin{eqnarray}\label{xp}
x_i(\bw, \bq) = \left\{\begin{array}{ll} 1 & \mbox{if }i =
\mbox{argmax}_k \left\{q_{k} M_k'(w_k)\right\} \\ 0 &
\mbox{otherwise.}\end{array}\right.
\end{eqnarray}
Here, ties among $q_kM_k'(w_k)$ are broken arbitrarily. For the
$(\epsilon n + 1)$th to the $n$th arrival, the allocation rule
$x_{ij} = x_i(\hat{\bu}, \bdb_j)$ is used.
\end{enumerate}
\end{algorithm}

Now we provide some intuition for the algorithm. The idea of the
algorithm is to use the first $\epsilon n$ inputs to learn an
approximate $\hat{\bu}$ and then use it to make all the future
allocations based on the complementarity conditions between the
primal and dual problems ((\ref{primal}) and (\ref{dual})). Here
$\hat{\bu}$ is solved from ($\bf P_{\epsilon}$) which projects the
allocation in the first $\epsilon n$ inputs to the entire problem.
The decision rule in (\ref{xp}) can be explained as choosing the $i$
with the highest product of the nominal bid value $b_{ij}$ and the
marginal contribution rate to the total projected reward
$M'_i(\hat{\bu})$. Note that a similar idea has been used to
construct algorithms for an online matching problem with linear
objective functions (see e.g., \citealt{Devanur, wang,
Molinaro2012}). However, the analyses of those algorithms all depend
on the linearity of the objective function which we do not possess
in this problem. Instead, an analysis with the use of concavity is
required in our analysis, making it quite different from those in
the prior literature. In the following, we assume without loss of
generality that $\max_{i,j} b_{ij} \le 1$ (we can always scale the
inputs to make this hold). We also make a technical assumption as
follows:

\begin{assumption}\label{assumption:general}
The inputs of the problem are in a {\emph general position}. That is,
for any vector $\bp = (p_1,...,p_m) \neq 0$, there are at most $m$
terms among $\mbox{arg}\max_i \{b_{ij}p_i\}$, $j=1,...,n$, that are
not singleton sets.
\end{assumption}
The assumption says that we only need to break ties in (\ref{xp}) no
more than $m$ times. This assumption is not necessarily true for all
inputs. However, as pointed out by \citet{Devanur} and \citet{wang},
one can always perturb $b_{ij}$ by adding a random variable
$\eta_{ij}$ taking uniform distribution on $[0, \eta]$ for some very
small $\eta$. By doing so the assumption holds with probability one
and the effect to the solution can be made arbitrarily small. Given
this assumption and by the complementarity conditions, we have the
following lemma, whose proof is in the Appendix.
\begin{lemma}\label{lem:dualsufficient}
\[
\epsilon\hat{u}_i - m \le \sum_{j=1}^{\epsilon n} b_{ij}
x_{i}(\hat{\bu}, \bdb_j) \le \epsilon\hat{u}_i + m.
\]
\end{lemma}
We first prove the following proposition about the performance of
the OLA, which relies on a condition of the solution to ($\bf
{P}_\epsilon$).

\begin{proposition}\label{prop:ola_condition1}
For any given $\epsilon \in (0, 1/2)$, if $\min_i\hat{u}_i\ge \Omega
\left( \frac{m\log{(m^2 n/\epsilon)}}{\epsilon^3} \right)$, then the
OLA is a $1-\epsilon$-competitive algorithm.
\end{proposition}

Before we prove Proposition \ref{prop:ola_condition1}, we define
some notation.
\begin{itemize}
\item We define the optimal offline solution to (\ref{primal}) by $(\bx^*, \bu^*)$ with optimal value $\OPT$.
\item Define $\sum_{j=1}^{n} b_{ij} x_{i}(\hat{\bu}, \bdb_j) =
\bar{u}_i$, note that $\bar{u}_i$ normally does not equal
$\hat{u}_i$.
\end{itemize}

We show the following lemma:
\begin{lemma}\label{lem:errorbound}
For any given $\epsilon \in (0, 1/2)$, if $\min_i\hat{u}_i\ge
\frac{12m\log{(m^2 n/\epsilon)}}{\epsilon^3} $,
then with probability $1-\epsilon$,
\begin{eqnarray}\label{ola_range}
(1-\epsilon)\hat{u}_i \le \bar{u}_i \le (1+\epsilon)\hat{u}_i,
\quad\mbox{for all }i.
\end{eqnarray}
\end{lemma}

{\noindent\bf Proof.}  The proof will proceed as follows: For any
fixed  $\hat{\bu}$, we define that a random sample (the first
$\epsilon n$ arrivals) $S$ is \textit{bad} for this $\hat{\bu}$ if
and only if $\hat{\bu} $ is the optimal solution to (\ref{partial})
for this $S$, but $ \bar{u}_i < \left( 1 - \epsilon \right)
\hat{u}_i$, or $\bar{u}_i > \left( 1 + \epsilon \right) \hat{u}_i$,
for some $i$. First, we show that the probability of a bad
sample is small for every fixed $\hat{\bu}$ (satisfying
$\min_i\hat{u}_i\ge \frac{12m\log{(m^2 n/\epsilon)}}{\epsilon^3}$)
and $i$. Then, we take a union bound over all distinct
 $i$ and $\hat{u}_i$s to prove the lemma.

To start with, we fix $\hat{\bu}$ and $i$. Define $Y_j = b_{ij}
x_{i} (\hat{\bu}, \bdb_j)$. By Lemma \ref{lem:dualsufficient} and
the condition on $\hat{u}_i$, we have
\[
(1-\epsilon^2) \epsilon \hat{u}_i \le \epsilon \hat{u}_i - m \le
\sum_{j\in S} Y_j \le \epsilon \hat{u}_i + m \le (1 + \epsilon^2)
\epsilon\hat{u}_i.
\]
Therefore, the probability of bad $S$
is bounded by the sum of the following two terms ($N = \{1,2,...,n\}$):
\begin{eqnarray} \label{sumbound}
P\left( \sum_{j \in S} Y_j \le \epsilon (1 + \epsilon^2) \hat{u}_i,
\sum_{j \in N} Y_j
> (1+\epsilon) \hat{u}_i \right) +  P\left( \sum_{j \in S} Y_j \ge
\epsilon (1-\epsilon^2) \hat{u}_i, \sum_{j \in N} Y_j < (1-\epsilon)
\hat{u}_i \right).
\end{eqnarray}
For the first term, we first define $Z_t =
\frac{(1+\epsilon)\hat{u}_i Y_t} {\sum_{j\in N} Y_j}$ and we have
\begin{eqnarray*}
P\left( \sum_{j \in S} Y_j \le \epsilon (1 + \epsilon^2) \hat{u}_i,
\sum_{j \in N} Y_j
> (1+\epsilon) \hat{u}_i \right) \le P \left(\sum_{j\in S} Z_j \le
\epsilon (1+ \epsilon^2) \hat{u}_i, \sum_{j\in N} Z_j =
(1+\epsilon)\hat{u}_i\right).
\end{eqnarray*}
Then we have
\begin{align*}
P \left( \sum_{j \in S} Z_j \le \epsilon (1 + \epsilon^2) \hat{u}_i,
\sum_{j \in N} Z_j = (1+\epsilon) \hat{u}_i \right) \le & P\left( |
\sum_{j \in S} Z_j - \epsilon \sum_{j \in N} Z_j |
> \frac{{\epsilon}^2}{2} \hat{u}_i ,
\sum_{j \in N} Z_j  =(1+\epsilon) \hat{u}_i\right) \nonumber \\
 \le & P\left( | \sum_{j \in S} Z_j - \epsilon \sum_{j \in N}
Z_j|
> \frac{{\epsilon}^2}{2} \hat{u}_i \left|
\sum_{j \in N} Z_j  =(1+\epsilon) \hat{u}_i\right.\right) \nonumber \\
\le& 2\exp \left( - \frac {\epsilon ^3 \hat{u}_i} {4(2+ \epsilon)}
\right) \le \frac{\epsilon}{2m(m^2n)^m} \doteq \delta.
\end{align*}
Here the second inequality follows from the Hoeffding-Bernstein's
inequality for sampling without replacement, see Lemma \ref{HB} in
the Appendix. Similarly, we can get the same result for the second
term in (\ref{sumbound}), which is also bounded by $\delta$.
Therefore, the probability of a bad sample is bounded by $2 \delta$
for fixed $\hat{\bu}$ and $i$.

Next, we take a union bound over all distinct $\hat{\bu}$s.
We call $\hat{\bu}$ and $\hat{\bu}'$ \textit{distinct} if and only
if they result in different allocations, i.e., $x_i(\hat{\bu},
\bdb_j) \neq x_i (\hat{\bu}', \bdb_j) $ for some $i$, $j$. Denote
$M_i'(\hat{u}_i) = v_i$. For each $j$, by the definition in
(\ref{xp}), the allocation is uniquely defined by the signs of the
following terms:
\begin{eqnarray*}
b_{ij}v_i - b_{i'j}v_{i'}, \quad \forall  1\le i < i' \le m.
\end{eqnarray*}
There are $m(m-1)/2$ such terms for each $j$. Therefore, the entire
allocation profiles for all the $n$ arrivals can be determined by
the signs of no more than $m^2 n$ differences. Now we find out how
many different allocation profiles can arise by choosing different
$v$s. By \citet{Orlik}, the total number of different profiles for
the $m^2n$ differences can not exceed $\left(m^2 n \right)^m$.
Therefore, the number of distinct $\hat{\bu}$s is no more than
$\left(m^2 n \right)^m$. Now we take a union bound over all
distinct $\hat{\bu}$s and $i = 1, \dots , m$, and Lemma
\ref{lem:errorbound}
follows. \hfill $\Box$\\

Next we show that the OLA achieves a near-optimal solution under the
condition in Proposition \ref{prop:ola_condition1}. We first
construct a feasible solution to (\ref{dual}):
\begin{eqnarray*}
\hat{v}_i = \hat{u}_i,\quad\quad \hat{y}_j =
\max_i\{b_{ij}M_i'(\hat{u}_i)\}.
\end{eqnarray*}
By Lemma \ref{lem:strongduality}, $\sum_{j=1}^n \hat{y}_j  +
\sum_{i=1}^m \left(M_i(\hat{u}_i)-M_i'(\hat{u}_i)\hat{u}_i\right)$
is an upper bound of $\OPT$. Thus, we have
\begin{align*}\label{deduction1}
 \OPT - \sum_{i=1}^m M_i(\bar{u}_i)
 \le & \sum_{i=1}^m
\left(M_i(\hat{u}_i) - \hat{u}_i M_i'(\hat{u}_i)\right) -
\sum_{i=1}^m M_i(\bar{u}_i) + \sum_{j=1}^n
\hat{y}_j\nonumber\\
 = & \sum_{i=1}^m (M_i(\hat{u}_i) -M_i(\bar{u}_i)) +
\sum_{i=1}^m\left(\bar{u}_iM_i'(\hat{u}_i) - \hat{u}_i M_i' (
\hat{u}_i) \right) - \sum_{i=1}^m\bar{u}_iM_i'(\hat{u}_i)
+\sum_{j=1}^n
\hat{y}_j\nonumber\\
 = & \sum_{i=1}^m \left(M_i(\hat{u}_i) -M_i(\bar{u}_i) +
(\bar{u}_i-\hat{u}_i)M_i'(\hat{u}_i)\right),
\end{align*}
where the last equality is because by the allocation rule
(\ref{xp}):
\begin{eqnarray*}
\sum_{j=1}^n \hat{y}_j = \sum_{i=1}^m\sum_{j=1}^n x_{i}(\hat{\bu},
\bdb_{j}) b_{ij} M_i'(\hat{u}_i) = \sum_{i=1}^m \bar{u}_i
M_i'(\hat{u}_i).
\end{eqnarray*}

Now, we claim that if condition (\ref{ola_range}) holds,
\begin{eqnarray*}
\sum_{i=1}^m \left(M_i(\hat{u}_i) -M_i(\bar{u}_i) +
(\bar{u}_i-\hat{u}_i)M_i'(\hat{u}_i)\right)\le 2\epsilon
\sum_{i=1}^m M_i(\bar{u}_i).
\end{eqnarray*}
We consider the following two cases:
\begin{itemize}
\item Case 1:  $\bar{u}_i \le \hat{u}_i $. In this case,
\begin{align*}
M_i(\hat{u}_i) -M_i(\bar{u}_i) +
(\bar{u}_i-\hat{u}_i)M_i'(\hat{u}_i)
 \le M_i(\hat{u}_i) - M_i(\bar{u}_i) \le \left| \frac{\hat{u}_i -
\bar{u}_i} {\bar{u}_i} \right| M_i(\bar{u}_i)\le 2\epsilon
M_i(\bar{u}_i),
\end{align*}
where the second inequality holds because of the concavity of
$M_i(\cdot)$.
\item Case 2: $\bar{u}_i > \hat{u}_i $. In this case,
\begin{align*}
 M_i(\hat{u}_i) -M_i(\bar{u}_i) +
(\bar{u}_i-\hat{u}_i)M_i'(\hat{u}_i)
 \le (\bar{u}_i-\hat{u}_i)M_i'(\hat{u}_i) \le
\left|\frac{\bar{u}_i-\hat{u}_i} {\hat{u}_i}\right| M_i(\hat{u}_i)
\le  \epsilon M_i(\bar{u}_i).
\end{align*}
Again, the second inequality is due to the concavity of
$M_i(\cdot)$.
\end{itemize}

Thus, under the condition that $\min_i \hat{u}_i \ge
\frac{12m\log{(m^2 n/\epsilon)}}{\epsilon^3}$, with probability
$1-\epsilon$,
\begin{eqnarray*} \label{Optim}
\OPT - \sum_{i=1}^m M_i(\bar{u}_i)
 ~\le~ 2 \epsilon \sum_{i=1}^m M_i(\bar{u}_i) ~\le ~ 2\epsilon \OPT,
\end{eqnarray*}
i.e., $\sum_{i=1}^m M_i(\bar{u}_i)\ge (1-2\epsilon)\OPT$.

Lastly, we note that the actual allocation in our algorithm for $i$
is $\tilde{u}_i = \sum_{j=\epsilon n + 1}^{n} b_{ij}
x_{i}(\hat{\bu}, \bdb_j)$ (since we ignore the first $\epsilon n$
arrivals). By Lemma \ref{lem:dualsufficient}, we have
\begin{eqnarray*}
\tilde{u}_i = \bar{u}_i - \sum_{j=1}^{\epsilon n}
b_{ij}x_{i}(\hat{\bu},\bdb_j) \ge \bar{u}_i -
\epsilon(1+\epsilon^2)\hat{u}_i.
\end{eqnarray*}
Thus when condition (\ref{ola_range}) holds, $ \tilde{u}_i\ge
(1-3\epsilon)\bar{u}_i$. Therefore,
\begin{eqnarray*}
\sum_{i=1}^{m} M_i(\tilde{\bu}) \ge \sum_{i=1}^{m} M_i((1-
3\epsilon) \bar{u}_i)  \ge (1-3\epsilon)\sum_{i=1}^{m}
M_i(\bar{u}_i).
\end{eqnarray*}
The last inequality is due to the concavity of $M_i(\cdot)$s
and that $M_i(0) =0$. Therefore, given $\min_i \hat{u}_i \ge
\frac{12m\log{(m^2 n/\epsilon)}}{\epsilon^3}$, with probability
$1-\epsilon$,
\begin{eqnarray*}
\sum_{i=1}^{m} M_i(\tilde{u}_i) \ge (1-5\epsilon) \OPT.
\end{eqnarray*}
Therefore, Proposition \ref{prop:ola_condition1} is proved. \hfill
$\Box$

Proposition \ref{prop:ola_condition1} shows that the OLA is
near-optimal under some conditions on $\hat{\bu}$. However,
$\hat{\bu}$ is essentially an output of the algorithm. Although such
types of conditions are not uncommon in the study of online
algorithms (e.g., in the result of \citealt{Devanur,
feldman2010online}), it is quite undesirable. In the following, we
address this problem by providing a set of sufficient conditions
which only depend on the input parameters (i.e., $m$, $n$, $\bdb$s
and $M(\cdot)$s). We show that our algorithm achieves near-optimal
performance under these conditions. We start with the following
lemma.

\begin{lemma}\label{lem:conditionforu}
For any $C>0$, suppose the following condition holds:
\begin{eqnarray}\label{condition_ola}
n \ge \max\left\{\frac{12 \log{(m/\epsilon)}}{\epsilon \bar{b}^2},
\frac{4mCF(M,\eta)}{\epsilon\bar{b}}\right\}
\end{eqnarray}
where $\bar{b} = \frac{1}{n}\min_{i}\{\sum_{j=1}^n b_{ij}\}$, $\eta
= \min_{i,j}\{b_{ij}|b_{ij} > 0\}$ and $F(M,\eta)$ is such that
\begin{eqnarray*}
M_i'(\eta F(M,\eta)C) < \eta M'_{i'}(C), \forall i, i'.
\end{eqnarray*}
Then with probability $1-\epsilon$, $\hat{u}_i \ge C$, for all $i$.
\end{lemma}
The proof of Lemma \ref{lem:conditionforu} is relegated to the
Appendix (it is proved together with Lemma
\ref{lem:condition_dynamic}). Now combining Proposition
\ref{prop:ola_condition1} and Lemma \ref{lem:conditionforu}, we have
the following result for the OLA:
\begin{proposition}\label{prop:1}
Fix any $\epsilon\in(0,1/2)$. Suppose
\begin{eqnarray}\label{condition_ola2}
n \ge \max\left\{\frac{12 \log{(m/\epsilon)}}{\epsilon \bar{b}^2},
\frac{4mCF(M,\eta)}{\epsilon\bar{b}}\right\}
\end{eqnarray}
where $\bar{b} = \frac{1}{n}\min_{i}\{\sum_{j=1}^n b_{ij}\}$, $\eta
= \min_{i,j}\{b_{ij}|b_{ij} > 0\}$ and $F(M,\eta)$ is such that
\begin{eqnarray}\label{condition_ola3}
M_i'(\eta F(M,\eta)C) < \eta M'_{i'}(C), \forall i, i'
\end{eqnarray}
with $C = \frac{12m\log{(m^2n/\epsilon)}}{\epsilon^3}$. Then the OLA
is $1-\epsilon$-competitive under the random permutation model.
\end{proposition}
Here we give some comments on the definition of $F(M, \eta)$. The
definition of $F(M,\eta)$ basically ensures that we rule out the
possibility that one $i$ receives nearly all the allocation while
some others receive almost none. Note that such $F(M,\eta)$ always
exists and is finite if $\lim_{x\rightarrow \infty} M'_i(x) = 0$ for
all $i$. In practice, this is usually true as there is usually a
upper bound on the possible reward from each bidder $i$. In
particular, if $M_i(\cdot) = M(\cdot)$ for all $i$ and
$\lim_{x\rightarrow\infty}M'(x) = 0$, then one can choose
\begin{eqnarray*}
F(M,\eta) = \frac{{M'^{-1}(\eta M'(C))}}{\eta},
\end{eqnarray*}
where $M'^{-1}(\cdot)$ denotes the inverse function of $M'(\cdot)$.
For example, if one chooses $M_i(x) = x^p$ ($0<p<1$), then one can
further choose $F(M,\eta)\ge \frac{2}{\eta^{(2-p)/(1-p)}}$.
Therefore, in most practical situations, one can view $F(M,\eta)$ as
a constant. Finally, we want to remark that the conditions in Lemma
\ref{lem:conditionforu} (or Proposition \ref{prop:1}) are only one
set of sufficient conditions which have the nice feature of only
depending on the problem inputs. In practice, one can always resort
to the condition in Proposition \ref{prop:ola_condition1} ($\min_i
\hat{u}_i \ge \frac{12m\log{(m^2n/\epsilon)}}{\epsilon^3}$) if they
are more favorable. In addition, as we will show in our numerical
tests in Section \ref{sec:numerical}, our algorithm performs quite
well even if some of the conditions in Lemma \ref{lem:conditionforu}
are not satisfied. Therefore, the applicability of our algorithm
could be well beyond what the conditions require.

\section{Dynamic Learning Algorithm}
\label{sec:dla}

In the previous section, we introduced the OLA
that can achieve near-optimal performance. While the OLA illustrates
the ideas of our approach and requires solving a convex
optimization problem only once, the conditions it requires to achieve
near-optimality are stricter than what we claim in Theorem
\ref{thm:main}. In this section, we propose an enhanced algorithm
that lessens the conditions and thus improves the OLA.

The main idea for the enhancement is the following: In the one-time
learning algorithm, we only solve a partial optimization problem
once. However, it is possible that there is some error for that
solution due to the random order of arrival. If we could modify the
solution as we gather more data, we might be able to improve the
performance of the algorithm. In the following, we introduce a
dynamic learning algorithm based on this idea, which updates the
{\it allocation policy} every time the history doubles; that is, it
computes a new $\hat{\bu}$ at time $ t = \epsilon n, 2 \epsilon n, 4
\epsilon n, \dots$ and uses it to perform the matching for the next
time period. We define the
following problem: 
\begin{equation}
\begin{array}{rrlll}\label{DPpartial}
({\bf P_\ell}) & \mbox{maximize}_{\bx, \bu} & \sum_{i=1}^m M_i(u_i)\nonumber\\
& \mbox{s.t.} & \sum_{j=1}^{\ell} \frac{n}{\ell} b_{ij} x_{ij}
= u_i, & \forall i\nonumber\\
&& \sum_{i=1}^m x_{ij} \le 1, &  \forall j\nonumber\\ & & x_{ij} \ge
0, &\forall i,j.
\end{array}
\end{equation}
We further define $(\bx^\ell, \bu^\ell)$ to be the optimal solution to
(${\bf P_\ell}$).

We define the dynamic learning algorithm as follows:

\begin{algorithm}
\caption{Dynamic Learning Algorithm (DLA)}\label{alg:dynamic}
\begin{enumerate}
\item During the first $\epsilon n$ arrivals, no allocation is made.
\item For $r = 0,1,...$, for $2^rn\epsilon < j \le 2^{r+1}n\epsilon$, set $x_{ij} = x_i(\bu^\ell, \bdb_j)$ for all $i$,
where $\ell =  \lceil 2^rn\epsilon\rceil$ .
\end{enumerate}
\end{algorithm}

In the following, without loss of generality, we assume that $L =
-\log_2 {\epsilon}$ is an integer (otherwise one can just choose a
smaller $\epsilon$ and prove the same result). Define $\ell_k =
2^{k-1}\epsilon n $, $k = 1,...,L$, and define $\hat{\bu}^k =
\bu^{\ell_k}$. We first prove the following proposition:
\begin{proposition}\label{prop:dynamic}
If for all $k$, $\min_i \hat{u}^k_i\ge \Omega \left\{
\frac{m\log{(m^2n/\epsilon)}}{\epsilon^2} \right\}$, then the DLA is
$1-\epsilon$-competitive under the random permutation model.
\end{proposition}

Before we proceed to the proof, we first define some more notation.
We define: $$ \bar{u}_i^k = \sum_{j = \ell_k +1}^{\ell_{k+1}}
b_{ij}x_{i}(\hat{\bu}^k, \bdb_j),\quad \tilde{u}^k_i = \sum_{j=1}^n
b_{ij}x_{i}(\hat{\bu}^k,\bdb_j), \quad \bar{u}_i = \sum_{k=1}^L
\bar{u}_i^k.$$ Note that in these definitions, $\bar{u}_i^k$ is the
allocated values for $i$ in the period $\ell_{k}+1$ to $\ell_{k+1}$
using $\hat{\bu}^k$, which is the actual allocation in that period.
$\tilde{u}_i^k$ is the allocation for $i$ in all periods if
$\hat{\bu}^k$ is used. $\bar{u}_i$ is the actual allocation for
$i$ during the entire algorithm. We first prove the following lemma
bounding the differences between $\bar{u}_i^k$, $\tilde{u}_i^k$ and
$\hat{u}_i^k$.

\begin{lemma}\label{lem:dynamic}
If $\min_i \hat{u}_i^k \ge
\frac{16m\log{(m^2n/\epsilon)}}{\epsilon^2}$, then with probability
$1-\epsilon$, for all $i$,
\begin{eqnarray}\label{ubarbound}
\left(1-\epsilon\sqrt{\frac{n}{\ell_k}}\right)\hat{u}_i^k \le
\frac{n}{\ell_k} \bar{u}_i^k \le
\left(1+\epsilon\sqrt{\frac{n}{\ell_k}}\right)\hat{u}_i^k
\end{eqnarray}
and
\begin{eqnarray}\label{utildebound}
\left(1-\epsilon\sqrt{\frac{n}{\ell_k}}\right)\hat{u}_i^k \le
\tilde{u}_i^k \le
\left(1+\epsilon\sqrt{\frac{n}{\ell_k}}\right)\hat{u}_i^k.
\end{eqnarray}
\end{lemma}
Lemma \ref{lem:dynamic} shows that with high probability,
$\frac{n}{\ell_k}\bar{u}_i^k$, $\tilde{u}_i^k$ and $\hat{u}_i^k$ are
close to each other. In particular, when $k$ is small, the factor
$(1\pm \epsilon\sqrt{n/\ell_k})$ is relatively loose while as $k$
increases, the factor becomes tight. The proof of Lemma
\ref{lem:dynamic} is similar to that of Lemma \ref{lem:errorbound}
and is relegated to the Appendix.

The next lemma gives a bound on the revenue obtained by the DLA.
\begin{lemma}\label{lem:dynamic_revenue}
If $\hat{u}_i^k\ge \frac{16m\log{(m^2n/\epsilon)}}{\epsilon^2}$ for
all $i$ and $k$, then with probability $1-\epsilon$,
\begin{eqnarray*}\label{revenuebound1}
\sum_{i=1}^m M_i\left(\frac{n}{\ell_i}\bar{u}_i^k\right) \ge
\left(1-6\epsilon\sqrt{\frac{n}{l_k}}\right) \OPT.
\end{eqnarray*}
\end{lemma}
The proof of Lemma \ref{lem:dynamic_revenue} can be found in the
Appendix.

Finally, we prove Proposition \ref{prop:dynamic}. We bound the
objective value of the actual allocation. Note that the actual
allocation for each $i$ can be written as
\begin{eqnarray*}
\sum_{k=1}^L \bar{u}_i^k = \sum_{k=1}^L \alpha_k
\frac{n}{\ell_k}\bar{u}_i^k,
\end{eqnarray*}
where $\alpha_k = \frac{\ell_k}{n}$. By the property of concave
functions, we have
\begin{align*}
 \sum_{i=1}^m M_i\left(\sum_{k=1}^L\bar{u}_i^k\right)
= \sum_{i=1}^m M_i\left(\sum_{k=1}^L \alpha_k\frac{n}{\ell_k}
\bar{u}_i^k + \left(1-\sum_{k=1}^L \alpha_k\right) \cdot 0\right)
\ge  \sum_{i=1}^m\sum_{k=1}^L \alpha_k
M_i\left(\frac{n}{\ell_k}\bar{u}_i^k\right).
\end{align*}
By Lemma \ref{lem:dynamic_revenue}, with probability $1-\epsilon$
\begin{align*}
\sum_{i=1}^m\sum_{k=1}^L \alpha_k
M_i\left(\frac{n}{\ell_k}\bar{u}_i^k\right)  \ge  \sum_{k=1}^L
\frac{\ell_k}{n}\left(1 -6\epsilon\sqrt{\frac{n}{\ell_k}}\right)
\OPT
 = & (1-\epsilon) \OPT - 6\epsilon \sum_{k=1}^L
\sqrt{\frac{\ell_k}{n}} \OPT \\
 \ge & (1- 16\epsilon) \OPT,
\end{align*}
where the last inequality is because
\begin{eqnarray*}
\sum_{k=1}^L \sqrt{\frac{\ell_k}{n}} = \sqrt{\frac{1}{2}} +
\sqrt{\frac{1}{4}}... \le 1 +  \sqrt{2} \le 2.5.
\end{eqnarray*}
Therefore, Proposition \ref{prop:dynamic} is proved. \hfill $\Box$

Similar to Lemma \ref{lem:conditionforu}, we have the following
conditions on the input parameters such that with high probability,
the conditions in Proposition \ref{prop:dynamic} hold.

\begin{lemma}\label{lem:condition_dynamic}
For any $C>0$, suppose the following condition holds:
\begin{eqnarray}\label{condition_dla}
n \ge \max\left\{\frac{24 \log{(m/\epsilon)}}{\epsilon \bar{b}^2},
\frac{4mCF(M,\eta)}{\epsilon\bar{b}}\right\}
\end{eqnarray}
where $\bar{b} = \frac{1}{n}\min_{i}\{\sum_{j=1}^n b_{ij}\}$, $\eta
= \min_{i,j}\{b_{ij}|b_{ij} > 0\}$ and $F(M,\eta)$ is such that
\begin{eqnarray*}
M_i'(\eta F(M,\eta)C) < \eta M'_{i'}(C), \forall i, i'.
\end{eqnarray*}
Then with probability $1-\epsilon$, $\hat{u}_i^k \ge C$, for all
$i$.
\end{lemma}

The proof of Lemma \ref{lem:condition_dynamic} is given in the
Appendix. Finally, we combine Proposition \ref{prop:dynamic} and
Lemma \ref{lem:condition_dynamic}, and Theorem \ref{thm:main}
follows.

The same remark after Lemma \ref{lem:conditionforu} applies here. In
particular, the conditions in Lemma \ref{lem:condition_dynamic} is
only one set of sufficient conditions for our algorithm to achieve
the target performance. However, one may also use the conditions in
Proposition \ref{prop:dynamic} if they turn out to hold in practice.
In the next section, we show that the DLA works well even if the
conditions in Lemma \ref{lem:condition_dynamic} are not satisfied.

\section{Numerical Experiments}
\label{sec:numerical} In this section, we report some numerical test
results for our algorithms (both the OLA and the DLA). The objective
is to validate the strength of our approaches and investigate the
relationship between the performance of our algorithms and the input
parameters.

In our numerical tests, we consider the Adwords problem. We assume
there are $m$ advertisers (bidders), $n$ keywords arriving
sequentially, and $b_{ij}$ is the amount bidder $i$ would like to
pay to display his advertisement on keyword search $j$. We introduce
a base problem in which we set $m = 50$, $n = 10,000$ and $M_i(x) =
x^{p}$ with $p=0.9$. The bidding values $b_{ij}$ are generated in
the following way:
\begin{enumerate}
\item Assume there are $k = 100$ categories of keywords. For each
category $k$, there is a base valuation of bidder $i$, denoted by
$\bar{b}_{ik}$, which is generated according to the following distribution:
\begin{eqnarray*}
\bar{b}_{ik} = \left\{\begin{array}{ll} 0 & \mbox{with
probability } 0.7 \\
U[0.2,1] & \mbox{with probability } 0.3, \end{array}\right.
\end{eqnarray*}
where $U[a,b]$ denotes a uniformly distributed random variable
on $[a,b]$.
\item For each arriving keyword,
we first randomly choose a category. The probability for each
category $i$, denoted by $\rho_i$, is randomly chosen on the simplex
$\{\rho_i| \sum_{i=1}^k \rho_i = 1, \rho_i \ge 0 \}$. Then if
category $k_0$ is chosen, the final bid value for bidder $i$ will be
$\bar{b}_{ik_0} \cdot U[0.9, 1.1]$.
\end{enumerate}
Although the way $b_{ij}$ is chosen seems arbitrary, we believe it
reflects some major features of the bid values in practice. In the
Adwords problem, each bidder is interested in certain categories of
keywords. For example, a sport product company is interested in
keywords related to sports. The $\bar{b}_{ik}$s represent such
interest levels. Then the bidder $i$'s actual bid on such a keyword
is the base value $\bar{b}_{ik}$ multiplied by a random number,
which reflects some level of idiosyncrasies of each keyword arrival.
We also tested other ways to generate $b_{ij}$, and the test results
are similar. We will report those test results in the end of this
section.

To evaluate the performance of our algorithms, we introduce the
notion of Relative Loss (RL) defined as follows:
\begin{eqnarray*}
\mbox{RL} = 1 - \frac{\mbox{Actual Revenue}}{\mbox{Offline Optimal
Revenue}}.
\end{eqnarray*}

\begin{figure}
\centering
\includegraphics[width= 3in]{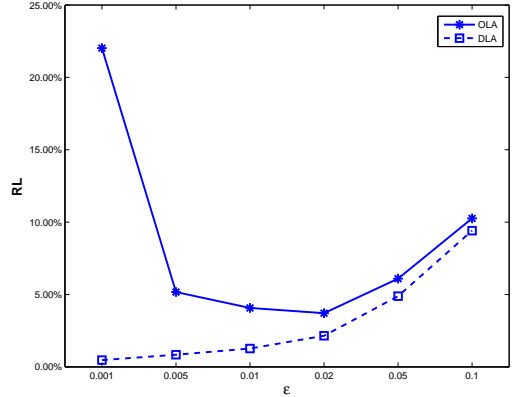}
\caption{Performance of the OLA and the DLA for Different Choices of
$\epsilon$}\label{fig:epsilon_line}
\end{figure}

In the numerical experiment, there is one key parameter we need to
set in both of our algorithms: $\epsilon$. In Theorem
\ref{thm:main} and Proposition \ref{prop:ola_condition1}, we gave
sufficient conditions on the inputs such that the algorithms will
have expected RL less than $\epsilon$. However, the theoretical
results are asymptotic and thus may not represent the best practical
choice of $\epsilon$. In Table \ref{table:testepsilon} and Figure 1,
we first test both our OLA and DLA with different choices of
$\epsilon$. We have the following observations from Table
\ref{table:testepsilon} and Figure \ref{fig:epsilon_line} (each
number in Table \ref{table:testepsilon} is the average of $100$
independent runs, the standard deviations of the results are
insignificant compared to the average value):

\begin{itemize}
\item For the DLA, choosing a smaller $\epsilon$ improves its
performance. There are two reasons for that. First, choosing a smaller
$\epsilon$ reduces the loss due to ignoring the first
$\epsilon n$ bids. Second, it increases the number of price updates
which help the decision maker to refine the decision
policy and achieve better performance. Therefore one should
choose a smaller $\epsilon$ in the DLA. 
\item For the OLA, the optimal choice of $\epsilon$ is more subtle.
There are two countervailing forces when one chooses a smaller
$\epsilon$. On one hand, by choosing a smaller $\epsilon$, the loss
due to the failure to allocate any bid during period $1$ to
$\epsilon n$ becomes smaller, which benefits the algorithm. On the
other hand, if $\epsilon$ is too small, the learned price may not be
accurate enough which may lead to poor allocation in the remaining
periods. In the test example, the optimal choice is $\epsilon =
0.02$.
\item The DLA outperforms the OLA for all choices of $\epsilon$.
\end{itemize}


\begin{table}
\centering
\begin{tabular}{|c||c|c|c|c|c|c|c|}
  \hline
  $\epsilon$ & 0.001 & 0.005 & 0.01 & 0.02 & 0.05 & 0.10 \\ \hline
  OLA   & 22.02\%  & 5.17\% & 4.07\% & 3.71\%  & 6.10\% & 10.26\%\\ \hline
  DLA   & 0.47\% & 0.84\% & 1.27\% & 2.15\% & 4.89\% & 9.41\% \\
  \hline
\end{tabular}
\caption{Performance of the OLA and the DLA for Different Choices of
$\epsilon$}\label{table:testepsilon}
\end{table}

\begin{figure}[ht]
\centering \subfigure[Different Problem Sizes]{
\includegraphics[width=3in]{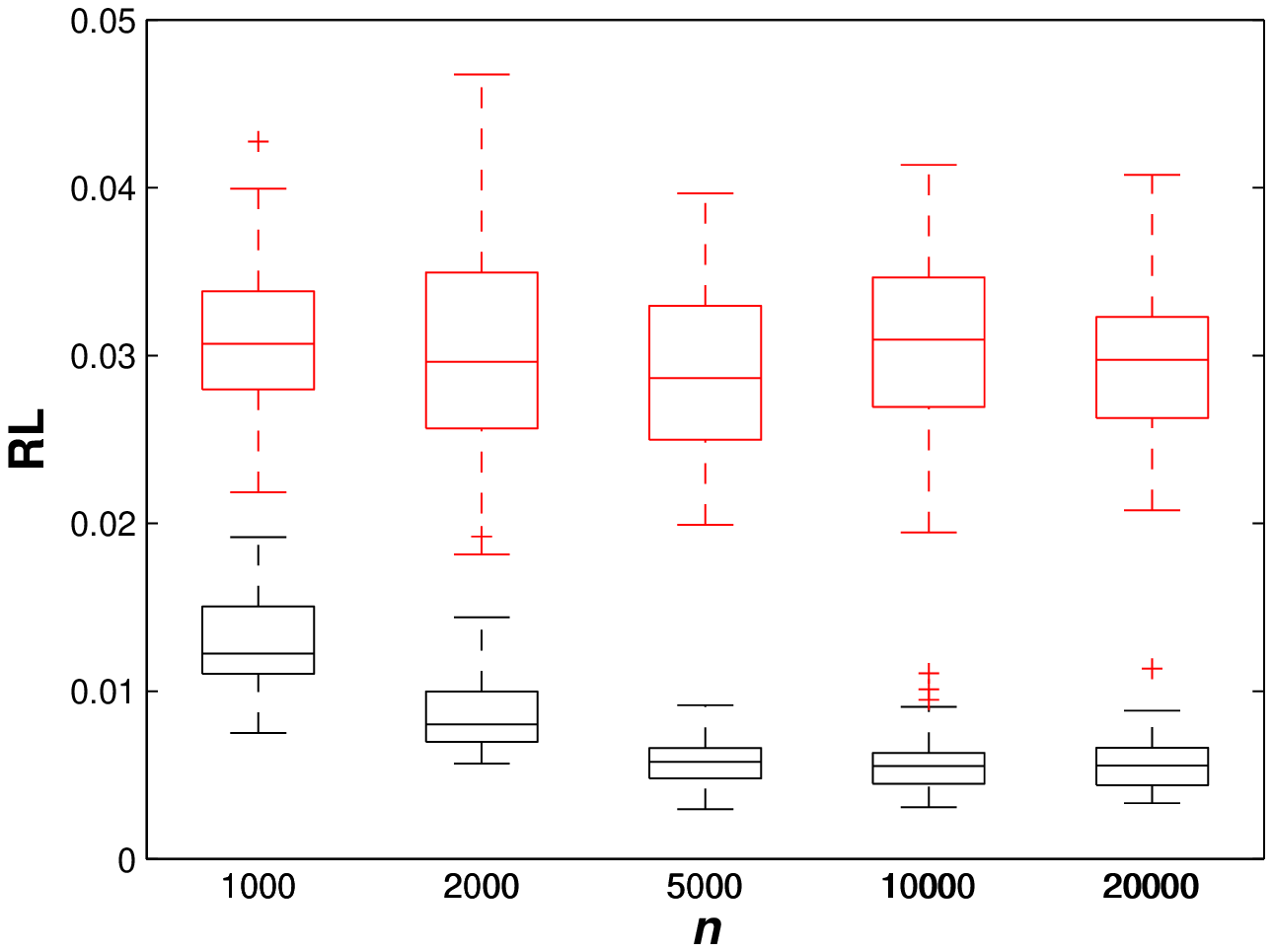}\label{n_line}
} \subfigure[Different $M(\cdot)$s]{
\includegraphics[width=3in]{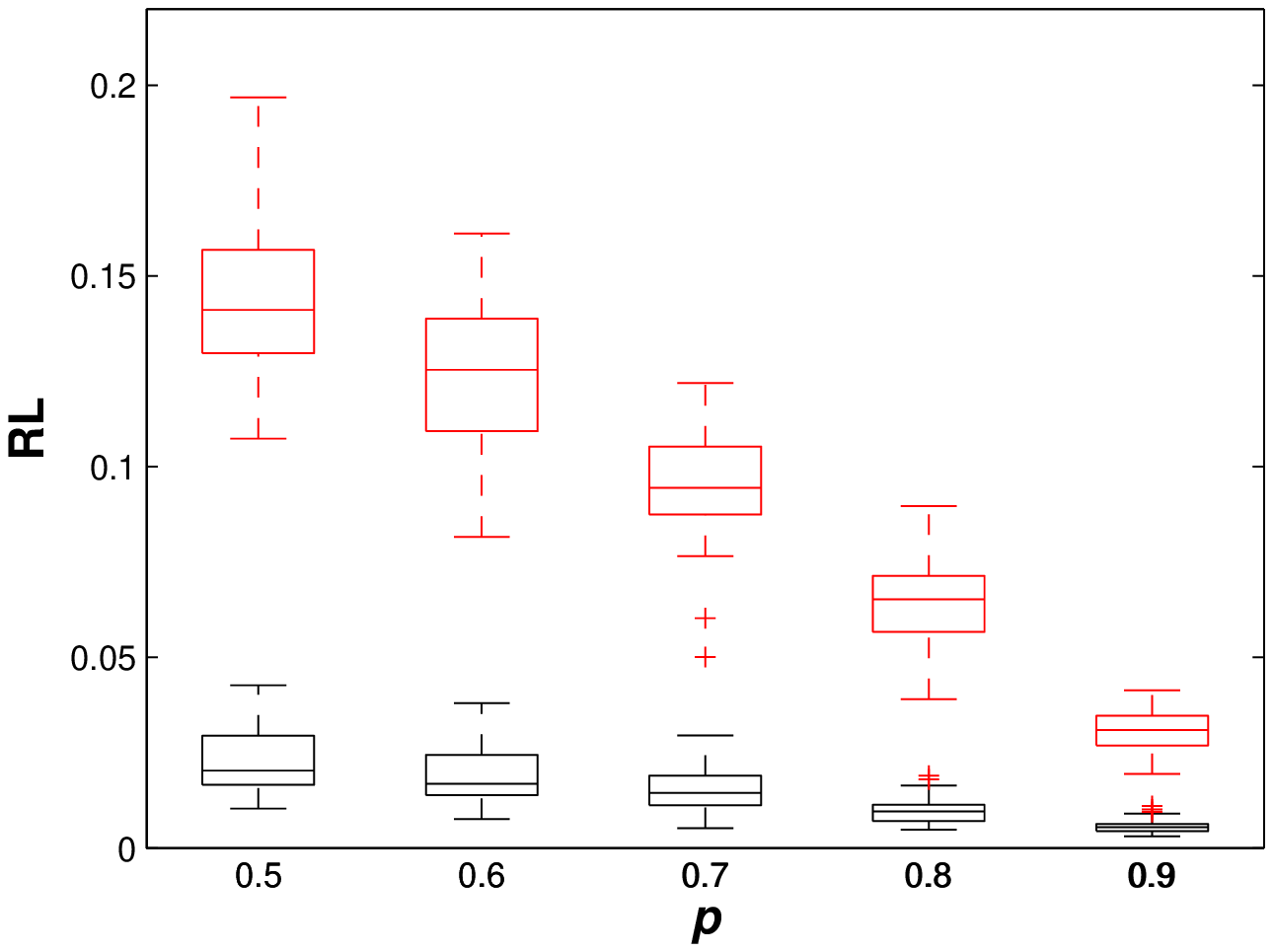}\label{p_line}
} \caption{Performance of the DLA (the Bottom Ones) and the Myopic
Policy (the Top Ones)} \label{fig:comparison}
\end{figure}

Next we focus ourselves on the DLA. As shown in Table
\ref{table:testepsilon}, we prefer to choose a smaller $\epsilon$ in
the DLA. In the following experiments, we will choose $\epsilon =
0.001$. Next we compare the performance of the DLA to a myopic
allocation method which simply allocates each incoming keyword to
the bidder with the highest $b_{ij}$ value. We also study the impact
of the two parameters, $n$ and $p$, on the performance of our
algorithm. We generate $100$ instances of the input $b_{ij}$ and
compare the average performance. The results of the average RL are
shown in Table \ref{table:testn} (the standard deviations are shown
in the parentheses) as well as in Figures \ref{n_line} and
\ref{p_line}.
\begin{table}[h]
\centering
\begin{tabular}{|c||c|c|c|c|c|}
  \hline
  $n$ & 1,000 & 2,000 & 5,000 & 10,000 & 20,000  \\ \hline
  DLA & 1.29\% (0.29\%) & 0.87\% (0.26\%)& 0.58\% (0.12\%)&
   0.57\% (0.17\%)& 0.57\% (0.16\%)\\ \hline
  Myopic & 3.07\% (0.47\%) & 3.03\% (0.61\%)& 2.90\% (0.49\%)&
   3.11\% (0.54\%)& 2.96\% (0.47\%)\\  \hline
  \hline
  $p$ & 0.5 & 0.6 & 0.7 & 0.8 & 0.9 \\ \hline
  DLA & 2.30\% (0.82\%) & 1.87\% (0.67\%)& 1.55\% (0.56\%) &
   0.99 \% (0.34\%)& 0.57\% (0.17\%)  \\
  \hline
  Myopic & 14.38\% (2.01\%)& 12.46\% (1.81\%)& 9.46\% (1.38\%)&
   6.51\% (1.05\%) & 3.11\% (0.54\%) \\
  \hline
\end{tabular}
\caption{Performance of the DLA and the Myopic
Policy}\label{table:testn}
\end{table}

From Table \ref{table:testn}, we can see that the DLA consistently
performs better than the myopic approach. In particular, the
performance of the DLA gradually improves when $n$ increases, while
the performance of the myopic approach seems to be insensitive to
the size $n$ of the problem. Moreover, even for small values of $n$,
the performance of the DLA is still very good. This means that the
DLA works well even for problems whose size is much smaller than
what Theorem \ref{thm:main} requires. For the parameter $p$, we can
see that both the DLA and the myopic algorithm deteriorate when $p$
decreases, but the DLA deteriorates much slower. Finally, we comment
that these results are computed when we ignore the first $\epsilon
n$ bids. In practice, one does not need to do that and the
performance of the DLA would be even better.

Finally, we repeat the above test for different setups of the
inputs. We fix the parameters $m = 50, n = 10,000$, and $M_i(x) =
x^p$ with $p=0.9$ in the base problem and generate $b_{ij}$s in the
following ways:
\begin{enumerate}
\item $b_{ij}$ follows a normal distribution (truncated at $0$ and $1$). The parameters of
each normal distribution (mean $\mu$ and standard deviation
$\sigma$) are randomly generated from a uniform distribution on $[0,
1]$.
\item $b_{ij}$ follows a Beta distribution. The parameters $(\alpha, \beta)$ of the Beta
distribution are generated from a uniform distribution on $[0,1]$.
\item $b_{ij}$ follows a
mixed normal and Beta distribution. That is, with probability $0.5$,
$b_{ij}$ follows a truncated normal distribution with mean $0.5$ and
standard deviation $0.5$, and with probability $0.5$, $b_{ij}$
follows a Beta distribution with $\alpha = \beta = 1/2$.
\end{enumerate}
Next, we compare the performance of the DLA (choose $\epsilon =
0.001$) and the myopic algorithm. For each case, we generate 100
instances of the input $b_{ij}$ and compare the average RL. The
results are shown in Table \ref{table:testn2}.

\begin{table}[h]
\centering
\begin{tabular}{|c|c|c|c|c|c|c|}
  \hline
  \multirow{6}{*}{Case 1} & $n$      & 1,000 & 2,000 & 5,000 & 10,000 & 20,000 \\ \cline{2-7}
                        & DLA    & 1.56\% (0.33\%)& 0.84\% (0.18\%)& 0.36\% (0.07\%)&
   0.21\% (0.05\%) & 0.14\% (0.03\%)  \\ \cline{2-7}
                        & Myopic & 1.45\% (0.45\%) & 1.37\% (0.47\%) & 1.42\% (0.45\%) &
   1.41\% (0.48\%) & 1.31\% (0.47\%) \\ \cline{2-7}
                        & $p$      & 0.5 & 0.6 & 0.7 & 0.8 & 0.9 \\ \cline{2-7}
                        & DLA  & 0.20\% (0.04\%) & 0.22\% (0.06\%) & 0.22\% (0.05\%) &
   0.21\% (0.03\%)& 0.21\% (0.05\%)   \\ \cline{2-7}
                        & Myopic & 9.10\% (2.01\%)& 7.98\% (2.00\%)& 5.71\% (1.32\%)&
   3.38\% (0.96\%) & 1.44\% (0.46\%) \\ \hline
   \hline

  \multirow{6}{*}{Case 2} & $n$      & 1,000 & 2,000 & 5,000 & 10,000 & 20,000 \\ \cline{2-7}
                        & DLA    & 1.19\% (0.24\%)& 0.61\% (0.12\%)& 0.25\% (0.08\%)&
   0.13\% (0.07\%) & 0.12\% (0.08\%)  \\ \cline{2-7}
                        & Myopic & 13.91\% (2.07\%) & 13.66\% (2.48\%) & 13.90\% (2.07\%) &
   13.78\% (1.97\%) & 13.62\% (2.12\%) \\ \cline{2-7}
                        & $p$      & 0.5 & 0.6 & 0.7 & 0.8 & 0.9 \\ \cline{2-7}
                        & DLA  & 0.20\% (0.16\%) & 0.29\% (0.20\%) & 0.21\% (0.12\%) &
   0.17\% (0.11\%)& 0.13\% (0.07\%)   \\ \cline{2-7}
                        & Myopic & 40.57\% (6.50\%)& 38.30\% (5.98\%)& 32.45\% (4.71\%)&
   24.67\% (3.26\%) & 13.46\% (2.07\%) \\ \hline
   \hline

   \multirow{6}{*}{Case 3} & $n$      & 1,000 & 2,000 & 5,000 & 10,000 & 20,000 \\ \cline{2-7}
                        & DLA    & 1.50\% (0.31\%)& 0.89\% (0.34\%)& 0.35\% (0.07\%)&
   0.21\% (0.06\%) & 0.14\% (0.01\%)  \\ \cline{2-7}
                        & Myopic & 10.84\% (2.50\%) & 11.22\% (2.64\%) & 11.30\% (2.44\%) &
   10.91\% (2.54\%) & 11.07\% (2.85\%) \\ \cline{2-7}
                        & $p$      & 0.5 & 0.6 & 0.7 & 0.8 & 0.9 \\ \cline{2-7}
                        & DLA  & 0.21\% (0.12\%) & 0.22\% (0.12\%) & 0.15\% (0.45\%) &
   0.15\% (0.35\%)& 0.21\% (0.06\%)   \\ \cline{2-7}
                        & Myopic & 37.00\% (7.43\%)& 34.02\% (6.82\%)& 30.22\% (5.89\%)&
   20.72\% (4.70\%) & 10.91\% (2.54\%) \\ \hline
\end{tabular}
\caption{Performance of the DLA and the Myopic Policy}
\label{table:testn2}
\end{table}

From Table \ref{table:testn2}, we can see that the DLA outperforms
the myopic approach under all the above three setups. The RL of the
DLA decreases as the problem size grows, while the RL of the myopic
policy is not sensitive to $n$. Also, as $p$ changes, the
performance of the DLA is rather stable, while the performance of
the myopic algorithm varies a lot. The overall trend of the DLA and
the myopic algorithm resembles that in the experiment in the
beginning of this section. Finally, we observe that the DLA seems
robust toward various problem setups, while the myopic approach does
not.

%


\section{Conclusion}
\label{sec:conclusion} In this paper, we propose a dynamic learning
algorithm for an online matching problem with concave returns. We
show that our algorithm achieves near-optimal performance when the
data arrives in a random order and satisfies some conditions. The
analysis is primal-dual based, however, the nonlinear objective
function requires us to work around nontrivial hurdles that do not
exist in previous work. Numerical experiment results show that our
algorithm works well in test problems.

\section{Acknowledgement}

We thank the anonymous referees for their insightful comments. The
research of both authors is supported by National Science Foundation
under research grant CMMI-1434541.

\bibliographystyle{ormsv080}
\bibliography{concave-mathor}

\newpage
\begin{APPENDIX}{}
In our proofs, we will frequently use the following
Hoeffding-Bernstein's Inequality for sampling without replacement:
\begin{lemma}[Theorem 2.14.19 in
\citealt{vandervaat}:] \label{HB} Let $u_1,u_2,...,u_r$ be random
samples without replacement from real numbers $\{c_1,c_2,...,c_R\}$.
Then for every $t>0$,
\begin{eqnarray*}
P\left(\left|\sum_{i=1}^r u_i-r\bar{c}\right|\ge t\right)\le
2\exp\left(-\frac{t^2}{2r\sigma_R^2+t\Delta_R}\right)
\end{eqnarray*}
where $\Delta_R=\max_{i}c_i-\min_{i}c_i$,  $\bar{c} =
\frac{1}{R}\sum_{i=1}^R c_i$, and $\sigma_R^2
=\frac{1}{R}\sum_{i=1}^R(c_i-\bar{c})^2$.\\
\end{lemma}

{\bf\noindent Proof of Lemma \ref{lem:strongduality}.} We first
write down the Lagrangian dual of (\ref{primal}). By associating
$p_i$ to the first set of constraints and $y_j$ to the second set of
constraints, the Lagrangian dual of (\ref{primal}) is:
\begin{equation}
\begin{array}{lll}\label{lagrangiandual} \mbox{inf}_{\bp, \by} & \sum_{j=1}^n y_j +
\sup_{u_i\ge 0}
\sum_{i=1}^m \left(M_i(u_i)-p_i u_i\right)\\
\mbox{s.t.} & y_j\ge b_{ij} p_i, \quad\quad \forall i\\
&y_j\ge 0,\quad\quad\forall j.
\end{array}
\end{equation}
Since the primal problem is convex and only has linear constraints,
Slater's condition holds, thus the strong duality theorem holds and
(\ref{primal}) and (\ref{lagrangiandual}) have the same optimal
value. Next we show that (\ref{dual}) and (\ref{lagrangiandual}) are
equivalent. To show this, assume the range of $M_i'(\cdot)$ on
$[0,\infty)$ is $(a_i,b_i]$ or $[a_i,b_i]$ (by the assumption that
$M(\cdot)$s are continuously differentiable, it must be either one
of these two forms). Now we argue that the optimal $p_i$ must be
within $[a_i, b_i]$ in (\ref{lagrangiandual}). First we must have
$p_i\ge a_i$, otherwise the term $\sup_{u_i\ge 0}
\left\{M_i(u_i)-p_i u_i\right\}$ goes to infinity as $u_i$ increases
and it cannot be the optimal solution to (\ref{lagrangiandual}). On
the other hand, if $p_i > b_i$, the optimal $u_i$ must be $0$, and
one can always set $p_i = b_i$ and achieves a smaller value of the
objective function. Therefore, $p_i \in [a_i,b_i]$ at optimality.

Now if $p_i \in (a_i, b_i]$ at optimality, one can always find one
$v_i$ such that $M_i'(v_i) = p_i$, and that $v_i$ must be the
optimal solution to $\sup_{u_i} \left\{M_i(u_i) - p_iu_i\right\}$
(the optimal solution must be attainable in this case). Therefore,
each feasible solution of (\ref{lagrangiandual}) will correspond to
a feasible solution of (\ref{dual}) and vice versa. The only case
left now is when $p_i = a_i$ at optimality. In this case,
$\sup_{u_i} \left\{M_i(u_i) - a_i u_i\right\} =
\lim_{x\rightarrow\infty} \left\{M_i(x) - a_ix\right\}$. Also, we
know that $\lim_{x\rightarrow\infty} M_i'(x) = a_i$, therefore,
there exists a sequence of feasible solution of (\ref{dual}) such
that the limit of the objective value equals the objective obtained
when $p_i =a_i$ in (\ref{lagrangiandual}). Therefore, the lemma is
proved. \hfill
$\Box$\\

{\bf\noindent Proof of Lemma \ref{lem:dualsufficient}.} To prove
this lemma, it suffices to show that for each fixed $i$,
$x_i(\hat{u}, \bdb_j)$ and $\hat{x}_{ij}$ (recall that $\hat{\bx} =
\{\hat{x}_{ij}\}$ is the optimal solution to (\ref{partial})) differ
by no more than $m$ terms. If this is true, then note that
$\sum_{j=1}^{\epsilon n} b_{ij}\hat{x}_{ij} = \epsilon\hat{u}_i$ and
$0\le b_{ij}\le 1$, the lemma holds.

To show that $x_i(\hat{u}, \bdb_j)$ and $\hat{x}_{ij}$ differ by no
more than $m$ terms, we first construct the dual problem of
(\ref{partial}) (according to (\ref{dual})):
\begin{equation*}
\begin{array}{lll}\label{dualappendix} \mbox{inf}_{\bv, \by} & \sum_{j=1}^{\epsilon n} y_j + \sum_{i=1}^m
\left(M_i(v_i) -
M_i'(v_i)v_i\right)\\
\mbox{s.t.}& y_j \ge \frac{b_{ij}}{\epsilon}M_i'(v_i),\quad\quad \forall i,j\\
& v_i\ge 0, \quad\quad\forall i\\
& y_j\ge 0, \quad\quad\forall j.
\end{array}
\end{equation*}
By Lemma 2.1, strong duality holds and thus any optimal solution
should satisfy the complementarity conditions. Among them we should
have $\hat{x}_{ij} ( y_j - \frac{b_{ij}}{\epsilon}M_i'(v_i)) = 0$.
Therefore, if there is no tie when we defined $x_i(\hat{u},
\bdb_j)$, we must have $x_i(\hat{u}, \bdb_j) = \hat{x}_{ij}$. By
Assumption \ref{assumption:general}, there are no more than $m$
ties. Thus, Lemma \ref{lem:dualsufficient} is proved. $\hfill\Box$\\

{\noindent\bf Proof of Lemma \ref{lem:dynamic}.}
\label{appendix:utildebound} We first prove (\ref{ubarbound}). The
idea is similar to the proof of the one-time learning case. For any
fixed $\hat{\bu}^k$, we define that a random sample $S$ (a sequence
of arrival) is \textit{bad} if and only if $\hat{\bu}^k$ is the
optimal solution to (${\bf P_{\ell_k}}$) but $\bar{\bu}^k$ does not
satisfy (\ref{ubarbound}) for some $i$. First, we show that the
probability of a bad sample is small for any fixed $\hat{\bu}^k$ and
fixed $i$. Then we take a union bound over all distinct
$\hat{\bu}^k$s and $i$s to show the result.

Fix $\hat{\bu}^k$ and $i$. We define $Y_j = b_{ij}x_{i}(\hat{\bu}^k,
\bdb_j)$. By Lemma \ref{lem:dualsufficient} and the assumption on
$\hat{u}_i^k$, we have
\begin{eqnarray*}
\frac{\ell_k}{n}\hat{u}_i^k - \epsilon^2 \hat{u}_i^k \le
\sum_{j=1}^{\ell_k} Y_j \le \frac{\ell_k}{n}\hat{u}_i^k + \epsilon^2
\hat{u}_i^k.
\end{eqnarray*}
Therefore, the probability of a bad sample is bounded by
the following two terms:
\begin{eqnarray}\label{tempterm1}
P\left(\sum_{j=1}^{\ell^k} Y_j \le \frac{\ell_k}{n}\hat{u}^k_i +
\epsilon^2 \hat{u}_i^k, \sum_{j=\ell_k+1}^{\ell_{k+1}} Y_j >
\frac{\ell_k}{n}(1+\sqrt{\frac{n}{\ell_k}}\epsilon)
\hat{u}^k_i\right) \nonumber\\
+ P\left(\sum_{j=1}^{\ell^k} Y_j \ge \frac{\ell_k}{n}\hat{u}^k_i -
\epsilon^2 \hat{u}_i^k, \sum_{j=\ell_k+1}^{\ell_{k+1}} Y_j <
\frac{\ell_k}{n}(1-\sqrt{\frac{n}{\ell_k}}\epsilon)
\hat{u}^k_i\right).
\end{eqnarray}

For the first term, we have
\begin{align*}
   & P\left(\sum_{j=1}^{\ell^k} Y_j \le  \frac{\ell_k}{n}\hat{u}^k_i + \epsilon^2\hat{u}_i^k,
\sum_{j=\ell_k+1}^{\ell_{k+1}} Y_j >
\frac{\ell_k}{n}\left(1+\sqrt{\frac{n}{\ell_k}}\epsilon\right)
\hat{u}^k_i\right) \\
  = & P\left(\sum_{j=1}^{\ell^k} Y_j \le \frac{\ell_k}{n}\hat{u}^k_i + \epsilon^2\hat{u}_i^k,
\sum_{j=1}^{\ell_{k+1}} Y_j >
\frac{\ell_k}{n}\left(2+\sqrt{\frac{n}{\ell_k}}\epsilon\right)
\hat{u}^k_i\right)\\
 \le &  P\left(|\sum_{j=1}^{\ell_k} Y_j - \frac{1}{2}
\sum_{j=1}^{\ell_{k+1}} Y_j| > \frac{\epsilon}{4}
\sqrt{\frac{n}{\ell_k}} \frac{\ell_k}{n}\hat{u}_i^k \right.
\left.\left|\sum_{j=1}^{\ell_{k+1}} Y_j
> \frac{\ell_k}{n}\left(2+\sqrt{\frac{n}{\ell_k}}\epsilon\right)
\hat{u}^k_i\right.\right)\\
 \le & 2\exp\left(-\frac{\epsilon^2 \hat{u}_i^k}{16}\right) \le  \frac{\epsilon}{2m(m^2n)^m} \doteq
 \delta.
\end{align*}
Here the second inequality follows from Lemma \ref{HB}, and the
third inequality is due to the condition of $\hat{u}_i^k$.
Similarly, we can get the bound for the second term in
(\ref{tempterm1}). Therefore, the probability of a bad sample is
bounded by $2\delta$.

Now we take union bound over all distinct $\hat{\bu}^k$ and
$i$. Similar to the proof of Lemma \ref{lem:errorbound}, we call
$\bu$s to be {\it distinct} if they result in different
allocations. As argued earlier, there are no more than $(m^2n)^m$
distinct $\bu$s. Therefore, we know that with probability
$1-\epsilon$, (\ref{ubarbound}) holds.

Next we prove (\ref{utildebound}). The idea is similar. Fix
$\hat{\bu}^k$ and $i$. We define $Y_j = b_{ij}x_{i}(\hat{\bu}^k,
\bdb_j)$. Applying Lemma \ref{HB}, we get
\begin{align*}
 & P\left(\sum_{j=1}^{\ell_k} Y_j \le \frac{\ell_k}{n} \hat{u}_i^k + \epsilon^2 \hat{u}_i^k,
\sum_{j=1}^n Y_j > \left(1 + \epsilon \sqrt{\frac{n}{\ell_k}}
\right)
\hat{u}_i^k \right) \\
\le & P\left(|\sum_{j=1}^{\ell_k} Y_j - \frac{\ell_k}{n}
\sum_{j=1}^n Y_j|
> \frac{\epsilon}{4}\sqrt{\frac{\ell_k}{n}} \hat{u}_i^k \right.
\left. \left| \sum_{j=1}^n Y_j > \left(1 +
\epsilon \sqrt{\frac{n}{\ell_k}}\right)\right. \hat{u}_i^k \right) \\
\le  & \exp\left(-\frac{\epsilon^2\hat{u}_i^k}{16}\right)  \doteq
\delta.
\end{align*}
Using the same argument as above, Lemma \ref{lem:dynamic} holds.
\hfill$\Box$ \\

{\noindent\bf Proof of Lemma \ref{lem:dynamic_revenue}.} The proof
consists of two main steps. First we show that with probability
$1-\epsilon$, the following is true for all $k$:
\begin{eqnarray}\label{revenuebound2}
\sum_{i=1}^m M_i(\tilde{u}_i^k) \ge
\left(1-2\epsilon\sqrt{\frac{n}{l_k}}\right) \OPT.
\end{eqnarray}
To show this, we follow a similar step when we prove the optimality
of the one-time learning algorithm. Define
\begin{eqnarray*}
\hat{v}_i^k  =  \hat{u}_i^k \mbox{ and } \hat{y}_i^k =  \max_{i}
\{b_{ij}M_i'(\hat{u}_i^k)\}.
\end{eqnarray*}
Since $(v_i^k, y_i^k)$ is a feasible solution to (\ref{dual}), we
know that
\begin{eqnarray*}
\sum_{j=1}^n \hat{y}_j^k + \sum_{i=1}^m (M_i(\hat{u}_i^k) -
M_i'(\hat{u}_i^k)\hat{u}_i^k)
\end{eqnarray*}
is an upper bound of $\OPT$. Therefore, by using the same argument
as in (12), we know that
\begin{align*}
\OPT - \sum_{i=1}^m M_i(\tilde{u}_i^k) \le
\sum_{i=1}^m(M_i(\hat{u}_i^k) - M_i(\tilde{u}_i^k) + (\tilde{u}_i^k
- \hat{u}_i^k) M_i'(\hat{u}_i^k)).
\end{align*}
Now for each term above, we consider two cases. If $\hat{u}_i^k \ge
\tilde{u}_i^k$, then
\begin{eqnarray*}
M_i(\hat{u}_i^k) - M_i(\tilde{u}_i^k) + (\tilde{u}_i^k -
\hat{u}_i^k) M_i'(\hat{u}_i^k) \le M_i(\hat{u}_i^k) -
M_i(\tilde{u}_i^k) \le
\frac{M_i(\tilde{u}_i^k)}{\tilde{u}_i^k}(\hat{u}_i^k
-\tilde{u}_i^k)\end{eqnarray*} and with probability $1-\epsilon$,
this is less than $
 2\epsilon \sqrt{\frac{n}{\ell_k}}M_i(\tilde{u}_i^k)$;
if $\hat{u}_i^k < \tilde{u}_i^k$, then
\begin{eqnarray*}
M_i(\hat{u}_i^k) - M_i(\tilde{u}_i^k) + (\tilde{u}_i^k -
\hat{u}_i^k) M_i'(\hat{u}_i^k)
 \le (\tilde{u}_i^k - \hat{u}_i^k)
M_i'(\hat{u}_i^k)  \le
\frac{M_i(\hat{u}_i^k)}{\hat{u}_i^k}(\tilde{u}_i^k - \hat{u}_i^k).
\end{eqnarray*}
Again, with probability $1-\epsilon$, this is less than $2\epsilon
\sqrt{\frac{n}{\ell_k}}M_i(\tilde{u}_i^k)$. Therefore, with
probability $1-\epsilon$,
\begin{eqnarray*}
\sum_{i=1}^m(M_i(\hat{u}_i^k) - M_i(\tilde{u}_i^k) + (\tilde{u}_i^k
- \hat{u}_i^k) M_i'(\hat{u}_i^k)) \le 2 \epsilon
\sqrt{\frac{n}{\ell_k}} \OPT.
\end{eqnarray*}
Therefore, (\ref{revenuebound2}) is proved. Next we show that
\begin{eqnarray*}
\sum_{i=1}^m M_i(\tilde{u}_i^k) - \sum_{i=1}^m
M_i(\frac{n}{\ell_k}\bar{u}_i^k) \le 4\epsilon
\sqrt{\frac{n}{\ell_k}}\OPT.
\end{eqnarray*}
To see this, by Lemma \ref{lem:dynamic}, we know that with
probability $1-\epsilon$,
\begin{eqnarray*}
\tilde{u}_i^k - \frac{n}{\ell_k}\bar{u}_i^k\le 2\epsilon
\sqrt{\frac{n}{\ell_k}} \hat{u}_i^k.
\end{eqnarray*}
Therefore, for each $i$, we have
\begin{eqnarray*}
\frac{\tilde{u}_i^k - \frac{n}{\ell_k} \bar{u}_i^k}{\tilde{u}_i^k}
\le \frac{2\epsilon \sqrt{\frac{n}{\ell_k}}
\hat{u}_i^k}{(1-\epsilon\sqrt{\frac{n}{\ell_k}})\hat{u}_i^k} \le 4
\epsilon\sqrt{\frac{n}{l_k}}.
\end{eqnarray*}
Now we analyze $M_i(\tilde{u}_i^k) -
M_i(\frac{n}{\ell_k}\bar{u}_i^k)$ for each $i$. We only need to
focus on the case when $\tilde{u}_i^k
> \frac{n}{\ell_k}\bar{u}_i^k$ (otherwise the difference is less than $0$).
In this case, by the concavity of $M_i(\cdot)$, we have
\begin{eqnarray*}
M_i(\tilde{u}_i^k) - M_i(\frac{n}{\ell_k}\bar{u}_i^k) \le
\frac{M_i(\tilde{u}_i^k)}{\tilde{u}_i^k} (\tilde{u}_i^k -
\frac{n}{\ell_k} \bar{u}_i^k) \le
4\epsilon\sqrt{\frac{n}{\ell_k}}M_i(\tilde{u}_i^k).
\end{eqnarray*}
Therefore, we have
\begin{align*}
\sum_{i=1}^m M_i(\tilde{u}_i^k) - \sum_{i=1}^m M_i(\frac{n}{\ell_k}
\bar{u}_i^k) \le  4\epsilon \sqrt{\frac{n}{\ell_k}} \sum_{i=1}^m
M_i(\tilde{u}_i^k) \le  4\epsilon \sqrt{\frac{n}{\ell_k}} \OPT.
\end{align*}
Together with (\ref{revenuebound2}), Lemma \ref{lem:dynamic_revenue}
holds. \hfill$\Box$ \\

{\bf\noindent Proof of Lemma \ref{lem:conditionforu} and Lemma
\ref{lem:condition_dynamic}.} First, we note that Lemma
\ref{lem:condition_dynamic} implies Lemma \ref{lem:conditionforu}
(except for the constant part, which can be strengthened easily by
only considering one $\ell_k$ in the following proof). Therefore, it
suffices to prove Lemma \ref{lem:condition_dynamic}.

We first prove for each $k$, with probability $1-\epsilon /
\log{(1/\epsilon)}$, $\min_i \hat{u}_i^k > C$. Then we take a union
bound to prove Lemma \ref{lem:condition_dynamic}. To show that for
each $k$, with probability $1-\epsilon / \log{(1/\epsilon)}$,
$\min_i \hat{u}_i^k > C$, first we show that with probability
$1-\epsilon / \log{(1/\epsilon)}$, $\sum_{j=1}^{\ell_k} b_{ij} \ge
\ell_k\bar{b}/2$ for all $i$. To see this, we use Lemma \ref{HB}, we
have for any $i$,
\begin{align*}
P\left(\left|\sum_{j=1}^{\ell_k} b_{ij} -
\frac{\ell_k}{n}\sum_{j=1}^nn_{ij}\right|\ge \ell_k\bar{b}/2\right)
\le 2 \exp(-\ell_k\bar{b}^2/12)
 < \frac{\epsilon}{m\log{(1/\epsilon)}},
\end{align*}
where the last inequality is due to condition (\ref{condition_dla}).
Next we show that given $\sum_{j=1}^{\ell_k} b_{ij} \ge
\ell\bar{b}/2$, there cannot exist an $i$ such that $\hat{u}_i^k <
C$ in the optimal solution to the partial program ($\bf
{P}_{\ell_k}$). We prove by contradiction. Let $K = \eta
\frac{\epsilon n \bar{b} - 2\epsilon C}{2mC}$. If there exists $i$
such that $\hat{u}_i^k < C$ in the optimal solution, then we argue
that there must exist $1\le j\le \ell_k$ such that
\begin{enumerate}
\item $j\in S_k = \{j: x_{ij} < 1, b_{ij} > \eta\}$, and
\item There exists $i'$ such that $x_{i'j} > 0$ and $\hat{u}_{i'}^k
\ge KC$.
\end{enumerate}%
Here these two conditions mean that there must exist a keyword $j$
such that we allocated it (at least partially) to bidder $i'$ whose
total allocation had already exceeded $KC$ when we could have
allocated it to bidder $i$ whose final allocation is less than $C$.

To see this, we note that we have proved with probability
$1-\epsilon / \log{(1/\epsilon)}$, $\sum_{j=1}^{\ell_k} b_{ij} \ge
\ell_k \bar{b}/2$. However, by the definition of $i$, $u_i^k < C$,
thus we also have $\sum_{j=1}^{\ell_k} b_{ij} x_{ij} \le \ell_k
C/n$. Therefore, combined with the assumption that $\max_{i,j}b_{ij}
\le 1$, there must exist at least $\ell_k\bar{b}/2 - \ell_kC/n$ $j$s
between $1$ and $\ell_k$ such that $x_{ij} < 1$ but $b_{ij} \ge
\eta$, i.e., $|S_k| \ge \ell_k \bar{b}/2 - \ell_kC/n$.

Next we show that among $j\in S_k$, there exists at least one $j$
such that $x_{i'j} > 0 $ while $\hat{u}_{i'}^k \ge KC$ for some
$i'$. We define $T_k = \{i: \hat{u}_i^k < KC\}$. We have
\begin{eqnarray}\label{temp4}
\sum_{i\in T,j\in S_k} x_{ij} \le \frac{1}{\eta} \sum_{i\in T,j\in
S_k} b_{ij}x_{ij} < \frac{mKC}{\eta}.
\end{eqnarray}
Here the second inequality is because $|T|< m$. However, we also
have
\begin{eqnarray}\label{temp5}
\sum_{i,j\in S_k} x_{ij} \ge \frac{\ell_k\bar{b}}{2}
-\frac{\ell_kC}{n}.
\end{eqnarray}
This is because $M_i(\cdot)$s are increasing, thus each $\sum_{i}
x_{ij}$ must equal $1$ at optimality. Then, by taking the difference
between (\ref{temp4}) and (\ref{temp5}), we have that
\begin{eqnarray*}
\sum_{i\not\in T, j\in S_k}x_{ij} >  \frac{\ell_k\bar{b}}{2} -
\frac{\ell_kC}{n} -\frac{mKC}{\eta} \ge 0.
\end{eqnarray*}
The last inequality is by the definition of $K$ and that $\ell_k\ge
\epsilon n$ for all $k$. Therefore, there exists $j\in S_k$ such
that the bid is allocated to some $i'$ with $\hat{u}_{i'}^k \ge KC$.
We denote such $j$ by $j^*$.

Finally, we consider another allocation that increases the
allocation of $j^*$ to $i$ while decreasing the allocation to $i'$.
The local change of the objective function at this point is:
\begin{eqnarray*}
M'_i(\hat{u}_i^k)b_{ij} - M'_{i'}(\hat{u}_{i'}^k)b_{i'j} \ge
M'_i(C)\eta - M'_{i'}(KC) > 0
\end{eqnarray*}
where the first inequality is due to the concavity of $M_i(\cdot)$s
and the last inequality is due to condition (\ref{condition_dla}).
However, this contradicts the assumption that the solution is
optimal. Thus, Lemma \ref{lem:condition_dynamic} is proved. \hfill
$\Box$

\end{APPENDIX}

\end{document}